\documentclass[a4paper,11pt]{article}
\usepackage{jcappub} 
\usepackage{lineno}
\usepackage{amsmath,amssymb, bm}
\usepackage{graphicx}
\usepackage{mathtools}
\usepackage{accents}

\arxivnumber{2511.13562} 
\title{Bimodular Gravity: Vacuum Evolution with a Frame-Dependent Phantom Crossing}

\author[a,b]{J. Hallam} 
\author[b]{and J. Magueijo}
\affiliation[a]{Institute of Cosmology and Gravitation, University of Portsmouth, \\
Burnaby Road, Portsmouth, PO1 3FX, United Kingdom}
\affiliation[b]{Abdus Salam Centre for Theoretical Physics, The Blackett Laboratory, \\
Imperial College London, Prince Consort Rd. London, SW7 2BZ, United Kingdom}

\emailAdd{james.hallam@port.ac.uk}

\abstract{Unimodular gravity recasts the cosmological constant as an integration constant, fixed by a constraint on the volume element rather than chosen in the action. We ask what becomes of this constant when matter couples not to the gravitational metric, but to a second metric disformally related to it through a scalar field. Imposing a volume constraint on each metric, we find a theory in which the scalar does not propagate, yet still drives a non-trivial expansion. Written as a single-metric theory, its kinetic term is fixed to a prescribed function of spacetime, and the cosmological constant is replaced by a vacuum contribution that need not be constant. Moreover, we find that this theory admits a phantom crossing through a purely frame-dependent mechanism. This construction, however, rests on a feature invisible with a single metric, and unimodular formalisms that are classically equivalent in that case cease to agree once there are two disformally related metrics. 
}

\begin{document}
\maketitle
\flushbottom

\newpage
\section{Introduction}
The late--time acceleration of the Universe has been well described, at the background level, by general relativity (GR) with a cosmological constant, which acts as a uniform source with an effective dark energy equation-of-state parameter $w=-1$ \cite{Riess_1998, Perlmutter_1999, Plank_2018}. 
However, recent observations have begun to place strain on this minimal picture, with DESI DR2 reporting mild tensions in $\Lambda$CDM and an improved fit for time-dependent dark energy when BAO data are combined with CMB and supernova measurements \cite{DESI_II}. More specifically, the preference is for an evolution in which the dark energy equation of state crosses the divide from a phantom to a quintessence behaviour at late times. Independently, the large discrepancy between quantum field theory estimates of the vacuum energy and the observed value of $\Lambda$ \cite{Zeldovich_1968, Weinberg_1989, Brout_2022} keeps open the possibility that cosmic acceleration is emergent or dynamical rather than a rigid constant, motivating theories that reconsider the provenance of $\Lambda$ and seek dynamical origins for the acceleration within an enlarged gravitational sector.

One such framework is that of unimodular gravity, which reframes the cosmological constant as an integration constant fixed by boundary conditions rather than a coupling in the action \cite{Anderson_1971,Alvarez_2013}. Whilst this does not, by itself, solve the cosmological constant problem \cite{2015-Bufalo}, it nevertheless provides a geometric origin for $\Lambda$, arising as an integration constant associated with enforcing a constraint on the metric determinant. 

Another alternative is scalar--tensor gravity \cite{Jordan_1959, BransDicke1961, Brans_1962}, in which a single spin–0 degree of freedom can modify the large-scale expansion whilst remaining compatible with local tests of gravity through screening mechanisms \cite{Sirera:2026klo}. The additional scalar may be further used to generalise matter couplings: matter fields interact with a matter metric $\hat g_{\mu\nu}$ that is related to $g_{\mu\nu}$ by a combination of conformal and disformal transformations mediated by a scalar field $\phi$ \cite{Bekenstein_1993, Gubitosi_2013EFTDE}. Such couplings are particularly appealing in cosmology, as the disformal component is typically suppressed in weak-field, quasi-static regimes but can become relevant on homogeneous backgrounds \cite{Koivisto_2008, BraxBurrage_2014,Sakstein_2014}, thereby providing a natural mechanism for dark energy.

Since unimodular gravity provides a geometric rationale for the cosmological constant, whilst scalar--tensor theories introduce additional degrees of freedom and generalised matter couplings, it is only natural to ask what becomes of this unimodular constant when matter sees a second metric. This work addresses precisely that question: given two metrics, the Einstein metric $g_{\mu\nu}$ governing the gravitational field equations and the matter metric $\hat g_{\mu\nu}$ to which non-gravitational fields couple, what is the analogue of the unimodular cosmological constant, and how does it affect the cosmological dynamics?

We show that the answer to this question depends on how unimodularity is implemented. In particular, imposing dual unimodular constraints on both $g_{\mu\nu}$ and $\hat g_{\mu\nu}$ breaks the classical equivalence between the different formalisms of unimodular gravity. Specifically, the nature of the resulting vacuum contribution, and the ability to achieve a phantom crossing, depend on how unimodularity is implemented.

The structure of this paper is as follows. In Sec.~\ref{sec:prelim} we review the relevant formulations of unimodular gravity and introduce the disformal bimetric framework used throughout. In Sec.~\ref{sec:bimod} we apply different unimodularisation procedures to this bimetric theory, comparing the resulting dynamics and clarifying the role played by the associated vacuum contributions. In Sec.~\ref{sec:cosmo} we specialise these constructions to homogeneous cosmology and examine their implications for the dark energy equation of state, including the possibility of phantom-divide crossing. In Sec.~\ref{sec:Covariant_Reformulation} we present a diffeomorphism-invariant reformulation which recovers the fixed-determinant construction in an appropriate gauge. Finally, we conclude in Sec.~\ref{sec:conclusion}.

Throughout this paper, we adopt the mostly-plus metric signature $(-+++)$, and define the canonical kinetic term of the scalar field as $X \coloneq -\tfrac12 g^{\mu\nu}\partial_\mu\phi\partial_\nu\phi$.
We work in natural units with $c=1$ and define the reduced Planck mass as $M_\text{Pl}^2 \equiv(8\pi G)^{-1 }$.

\section{Preliminaries}
\label{sec:prelim}

We begin by outlining the theoretical background required for the construction of bimodular gravity. Since the framework is centred on deriving the cosmological constant as a Lagrange multiplier, it is helpful to first recall how this arises in the different formulations of unimodular gravity. For brevity, we focus on the three versions reviewed extensively in \cite{2015-Bufalo}. We then turn to the scalar--tensor bimetric gravity theory of \cite{Clayton_2000, Clayton_2001, Moffat_2003, Magueijo_2003}, for it is the framework subject to unimodularisation in this work. Since this work is interested in purely the combined gravitational and biscalar effects, we neglect additional matter fields.

\subsection{Unimodular Gravity}
\label{sec:UG}
If one imposes a constraint on the metric via a Lagrange multiplier, that multiplier appears in the gravitational field equations and, for a suitable choice of constraint, can be identified with the cosmological constant. This modified formulation of GR is known as unimodular gravity, whose various formulations all yield field equations classically equivalent to GR$+\Lambda$. The differences amongst them lie only in how the determinant constraint (and thus the constancy of the associated Lagrange multiplier) is enforced, and hence in their off-shell symmetry content.

In the original fixed-determinant formulation of \textit{unimodular gravity} (UG), one constrains $\sqrt{-g}$ to a fixed scalar density $\varepsilon_0(x)$ by means of a Lagrange multiplier $\lambda$, 
\begin{equation}
    S_\text{UG} = \int d^4x \left[ \frac12 \sqrt{-g}R[g] - \lambda(x) \left(\sqrt{-g}-\varepsilon_0(x) \right)
    \right].
\end{equation}
It is the presence of the last term that breaks full diffeomorphism invariance. Variation with respect to $\lambda$ imposes the constraint $\sqrt{-g}=\varepsilon_0$, whilst variation with respect to $g_{\mu\nu}$ yields the Einstein equations
\begin{equation}
    G_{\mu\nu} +\lambda g_{\mu\nu} = 0.
    \label{eq:UG_EinsteinEquation}
\end{equation}
Taking the divergence of \eqref{eq:UG_EinsteinEquation}, and using the Bianchi identity $\nabla_\mu G^{\mu\nu}=0$, we find that the multiplier must satisfy $\nabla_\mu\lambda=0$. Thus, it is a derived result that $\lambda$ is constant on shell and can be identified with the cosmological constant $\Lambda \coloneq \lambda$ \cite{Dragon_1988, Dragon_1989, 1989-Unruh, Padilla_2015}. Hence the gravitational field equations of UG coincide with those of GR$+\Lambda$, with $\Lambda$ appearing as an integration constant rather than as a parameter in the action.

The covariant formulation of \textit{Henneaux and Teitelboim} (HT) \cite{1989-Henneaux} restores full diffeomorphism invariance by replacing the fixed-determinant constraint on the metric with a constraint on the divergence of an auxiliary vector density $\tau^\mu$, 
\begin{equation}
    S_\text{HT} = \int d^4x \left[
    \frac12 \sqrt{-g} R[g] - \lambda(x) \left( \sqrt{-g} - \partial_\mu\tau^\mu(x) \right)
    \right].
\end{equation}
Here, variation with respect to $\lambda$ enforces $\sqrt{-g}=\partial_\mu \tau^\mu$, whilst variation with respect to $\tau^\mu$ yields $\partial_\mu\lambda=0$, so the constancy of $\lambda$ is imposed already at the level of the action \cite{Jirousek_2023}. Variation with respect to $g_{\mu\nu}$ then gives the Einstein field equations with a more standard cosmological constant $\Lambda \coloneq \lambda$, so the HT formulation reproduces GR$+\Lambda$ whilst maintaining full diffeomorphism invariance. 

A closely related fully diffeomorphism-invariant formalism (DUG) trades the metric constraint for a divergence condition on an auxiliary vector $V^\mu$, 
\begin{equation}
    S_\text{DUG} = \int d^4x \left[
    \frac12\sqrt{-g} R[g] - \sqrt{-g}\left(\lambda(x)+V^\mu(x)\partial_\mu \lambda(x) \right)
    \right],
\end{equation}
which is equivalent to HT up to a boundary term. Variation of $V^\mu$ directly yields $\partial_\mu \lambda=0$, and variation with respect to $\lambda$ imposes $\nabla_\mu V^\mu =1$ \cite{2015-Bufalo}.

In all three cases, the classical field equations are on-shell equivalent to those of GR with a cosmological constant. Their differences are purely off shell, specifically in how the unimodular condition is implemented and how the constancy of $\lambda$ is imposed---derived via the Bianchi identity (UG) or assumed at the level of the action (HT/DUG). These distinctions are classically inconsequential in the single-metric setting but will become decisive for the failure of the classical equivalence once unimodularity is applied in a bimetric framework. To that end, the next section introduces the disformal bimetric theory that will subsequently be unimodularised.

\subsection{Bimetric Gravity}
In the bimetric gravity theory considered here, gravity is described by an \textit{Einstein metric} $g_{\mu\nu}$ and a \textit{matter metric} $\hat g_{\mu\nu}$, disformally related through a scalar field $\phi$, often referred to as the \textit{biscalar}, via
\begin{equation}
    \hat g_{\mu\nu} = g_{\mu\nu} +B(\phi) \partial_\mu\phi\partial_\nu\phi,
    \label{eq:disformal_relation}
\end{equation}
with the \textit{disformal coupling} $B(\phi)$. We restrict to the branch $B(\phi)>0$ and impose the condition $1-2BX>0$ which ensures invertibility of the disformal map and preservation of Lorentzian signature for timelike scalar gradients. In that case, from the Sherman–Morrison formula and matrix determinant lemma, the inverse and determinant of \eqref{eq:disformal_relation} follow as
\begin{align}
    \hat g^{\mu\nu} &= g^{\mu\nu} - \frac{B}{1-2BX}\partial^\mu\phi\partial^\nu\phi
    \label{eq:matter_inverse}
    , \\
    \sqrt{-\hat g} &= \sqrt{-g}\sqrt{1-2BX},
    \label{eq:matter_det}
\end{align}
so that the \textit{relative volume element} between the two metrics is
\begin{equation}
    \nu \coloneqq \frac{\sqrt{-\hat g}}{\sqrt{- g}}
    = \sqrt{1-2BX}.
    \label{eq:mathcalV}
\end{equation}
These are precisely the objects that will be subjected to unimodular constraints in Sec.~\ref{sec:bimod}. It may already be obvious how the different unimodular formalisms will affect $\nu$, and subsequently the biscalar dynamics.

For the biscalar to be dynamical, a kinetic term must be chosen. One may place $\phi$ in the Einstein frame, where it is canonical relative to $g_{\mu\nu}$ and luminal with respect to that frame, or in the matter frame, where conservation laws take their standard form \cite{2009-Magueijo}. For present purposes, it is more insightful to adopt the latter so that the gravitational sector alone sees two metrics. We thus consider the bimetric action 
\begin{equation}
    S =  \int  d^4x \left[
    \frac12\sqrt{-g}R 
    + \sqrt{-\hat g} 
    \hat X 
    \right],
    \label{eq:BGT}
\end{equation}
with $\hat X \coloneq -\frac12 \hat g^{\mu\nu}\partial_\mu\phi\partial_\nu\phi =X/\nu^2$ where the latter equality is a non-trivial result. Here, the matter stress tensor is conserved in the matter frame, $\hat \nabla_\mu \hat T^{\mu\nu}=0$, i.e. on the hatted scalar equations of motion. 

Indeed, by rewriting the theory in a single-metric presentation, obtained by direct substitution of \eqref{eq:matter_inverse}--\eqref{eq:mathcalV} into \eqref{eq:BGT}, one finds a particular k-essence theory,
\begin{equation}
S =  \int  d^4x \sqrt{-g} \left[ \frac12 R + K(\phi,X)  \right],
\qquad
K(\phi,X) = \frac{X}{\sqrt{1-2BX}} . 
\label{eq:BGT_kessence}
\end{equation}
This is a change of presentation, not of physical content. In the single-metric presentation, diffeomorphism invariance implies the usual on-shell conservation law $\nabla_\mu T^{\mu\nu}=0$. The two conservation laws are therefore not independent assumptions: they are equivalent on shell under the disformal field redefinition, although the stress tensors and covariant derivatives appearing in the two descriptions are different.\footnote{
Since, at fixed $\phi$, $\delta\hat g_{\mu\nu}=\delta g_{\mu\nu}$, one obtains
\begin{equation}
    \delta S_\phi 
    = \frac12 \int d^4x \sqrt{- \hat g}\, \left[ \hat T^{\mu\nu} \right]  \delta\hat g_{\mu\nu} 
    = \frac12 \int d^4x \sqrt{- g} \, \left[ \nu \hat T^{\mu\nu} \right]  \delta g_{\mu\nu}.
\end{equation}
Therefore $ T^{\mu\nu}  = \nu \hat T^{\mu\nu}$ for the contravariant stress tensors, with the important caveat that $\hat T^{\mu\nu}$ is raised using $\hat g^{\mu\nu}$, not $g^{\mu\nu}$. Equivalently, for the Einstein-frame covariant components, $T_{\mu\nu} = \nu g_{\mu\kappa}g_{\nu\lambda}\hat T^{\kappa\lambda}$.
}

If one were to include matter, its action should be coupled universally to a single physical metric,
\begin{equation}
    S_{\rm m} = S_{\rm m}[\tilde g_{\mu\nu}, \psi_{\rm m}],
    \qquad 
    \tilde g_{\mu\nu} = g_{\mu\nu}
    \quad\text{or}\quad
    \tilde g_{\mu\nu}= \hat g_{\mu\nu},
\end{equation}
in order to preserve the weak equivalence principle. The choice $\tilde g_{\mu\nu}=\hat g_{\mu\nu}$ is natural from the original bimetric perspective, since $\hat g_{\mu\nu}$ is then the matter metric. In this case, observables would then be defined in the matter frame.

\section{Unimodularising Bimetric Gravity}
\label{sec:bimod}

In this section, we subject the bimetric gravity action \eqref{eq:BGT} to the unimodularisation procedures of Sec.~\ref{sec:UG}. Whilst there is no classical difference between the two formalisms in the single metric case, we find here that that classical equivalence is broken when a disformal metric is present. In order to unimodularise the disformal bimetric setup in full generality, we impose one unimodular-type condition per metric using two independent Lagrange multipliers, $\lambda_1$ and $\lambda_2$, and two constraint functionals $\mathcal C_1[g]$ and $\mathcal C_2[\hat g]$. The general action is then
\begin{align}
    S =& \int d^4x \left[
    \frac12 \sqrt{-g} R[g]
    + \sqrt{-\hat g} \hat X
    -
    \lambda_1 \mathcal{C}_1[g]
    - \lambda_2 \mathcal{C}_2[\hat g] 
    \right].
    \label{eq:bimodular_gravity_action_general}
\end{align}
The choice of constraint functionals specifies the appropriate unimodular formalism, and thus the bimetric unimodular formalism, which we now present for the three approaches discussed in Sec.~\ref{sec:UG}. 

\subsection{Bimetric Unimodular Gravity Theory}
\label{subsec:biug}
We begin by subjecting \eqref{eq:bimodular_gravity_action_general} to the fixed-determinant constraints. Since $\nu$ is tied to the biscalar kinetic term, the constraints have direct consequences for the biscalar dynamics. Furthermore, in single-metric UG the constancy of the Lagrange multiplier, and hence its identification with $\Lambda$, was not assumed but derived from the Bianchi identity. We therefore close by computing the analogous relation to determine the nature of the bimetric multipliers.

\subsubsection{Fixed-Determinant Bimetric Constraints}
We thus define \textit{Bimetric Unimodular Gravity} (BUG) theory as
\begin{equation}
    S_{BUG} \supset -\int d^4x \left[
    \lambda_1\left( \sqrt{-g}-\varepsilon_1\right)
    +\lambda_2 \left( \sqrt{-\hat g} - \varepsilon_2 \right)
     \right],
    \label{eq:BUG_action}
\end{equation}
where $\varepsilon_{1,2}(x)$ are fixed scalar densities. Variation with respect to $\lambda_{1,2}$ enforces
\begin{equation}
    \sqrt{-g}=\varepsilon_1,\qquad \sqrt{-\hat g}=\varepsilon_2,
    \label{eq:BUG_constaints}
\end{equation}
Hence the relative volume element  \eqref{eq:mathcalV} is not dynamical, but fixed to the ratio of the fixed scalar densities
\begin{equation}
    \nu(x) = \frac{\varepsilon_2(x)}{\varepsilon_1(x)} = \sqrt{1-2B X}.
    \label{eq:BX-const}
\end{equation}
It follows that the biscalar kinetic density is kinematically locked to the fixed density ratio 
\begin{equation}
    X = \frac{1-\nu^2(x)}{2B(\phi)}.
    \label{eq:kin_lock}
\end{equation}
Differentiating \eqref{eq:kin_lock} gives
\begin{equation}
    \nabla_\mu X
    =
    -\frac{B_\phi X}{B}\nabla_\mu\phi
    -
    \frac{\nu}{B}\nabla_\mu\nu .
    \label{eq:app_gradX_BUG}
\end{equation}

Now, suppose that the scalar gradient is timelike, $X>0$. One may then define the unit four-velocity
\begin{equation}
    u_\mu=\frac{\nabla_\mu\phi}{\sqrt{2X}},
    \qquad
    u^\mu u_\mu=-1,
\end{equation}
together with the orthogonal projector
\begin{equation}
    h_{\mu\nu}=g_{\mu\nu}+u_\mu u_\nu .
\end{equation}
The four-acceleration of the scalar flow is
\begin{equation}
    a_\mu
    \coloneq
    u^\nu\nabla_\nu u_\mu
    =
    -\frac{1}{2X}h_\mu{}^\nu\nabla_\nu X .
\label{eq:app_acceleration_X}
\end{equation}
Thus the flow is geodesic if and only if the spatially projected gradient of $X$ vanishes, $h_\mu{}^\nu\nabla_\nu X=0 $. Projecting \eqref{eq:app_gradX_BUG} orthogonally to the scalar flow removes the first term, since it is parallel to $\nabla_\mu\phi$. Hence 
the acceleration becomes
\begin{equation}
    a_\mu
    =
    \frac{\nu}{2BX}
    h_\mu{}^\nu\nabla_\nu\nu 
    \neq0.
    \label{eq:app_BUG_acceleration}
\end{equation}
The same conclusion holds in the matter frame, since the corresponding acceleration is related by $a_\mu=\nu^2\hat a_\mu$. Generic BUG therefore does not classify as dusty dark energy \cite{Lim_2010}.

This also clarifies the status of the biscalar sound speed. Since the BUG scalar sector is kinematically constrained by the fixed relative volume element, it is not an ordinary freely propagating k-essence degree of freedom, and the standard k-essence expression for $c_s^2$ is not the relevant physical propagation speed. At the same time, the Dust of Dark Energy argument for an identically vanishing physical sound speed relies on geodesic flow \cite{Lim_2010}, or equivalently on the constraint reducing to $X=X(\phi)$. In this sense BUG is neither standard k-essence nor generically identical to Dust of Dark Energy, although it contains the latter as a special case if $\nu = \text{const}$ with $B(\phi)$. For the remainder of this work we take $B(\phi)=B$ to be constant and treat $\nu(x)$ as an arbitrary fixed function, since it is this combination that gives us the simplest novel behaviour. As such, the BUG constraint reduces to
\begin{equation}
    \nabla_\mu X
    =
    -\frac{\nu}{B}\nabla_\mu\nu .
    \label{eq:kin_lock_deriv}
\end{equation}
Interestingly, in the special case $\nu(x)=\nu$ is constant in addition to constant $B$, we mimic the constant kinetic constraint of Mimetic Gravity \cite{Chamseddine_2013, jirousek2022mimetickessence}.

But of course, no condition has yet been imposed upon $\lambda_{1,2}$. In single-metric UG, constancy of the multiplier is derived via the Bianchi identities, which then allows it to be identified with the cosmological constant. Therefore, the analogous relation must be derived through explicit calculation of the BUG Bianchi identity. For that, the metric and bimetric field equations must be determined.

\subsubsection{The Dynamics of BUG}
For a constant disformal coupling $B$, extremising the general BMG action \eqref{eq:bimodular_gravity_action_general} with respect to the biscalar gives the field equation\footnote{%
Since some components of $\hat\Box\phi$ are also constrained by~\eqref{eq:kin_lock_deriv}, the kinematic constraint must be treated consistently in order for the dynamics to remain self-consistent~\cite{Hallam_2024}.
}
\begin{align}
    \left[B (\hat X + \lambda_2) + 1 \right] \hat \Box \phi
    =
    - B\hat \nabla_\mu \hat X \hat \nabla^\mu \phi
    - B \hat \nabla_\mu \lambda_2 \hat \nabla^\mu \phi .
    \label{eq:BUG_ScalarEoM}
\end{align}
In BUG, however, $\hat X=X/\nu^2$ and its derivative are fixed by the kinematic constraints \eqref{eq:kin_lock} and \eqref{eq:kin_lock_deriv}. We keep \eqref{eq:BUG_ScalarEoM} in matter-frame notation for compactness.

Varying the BUG action \eqref{eq:BUG_action} with respect to the Einstein-frame metric, expressing all quantities in the Einstein frame, and dividing through by $\sqrt{-g}$ gives
\begin{align}
    G_{\mu\nu}+\lambda_1 g_{\mu\nu}
    =
    T_{\mu\nu},
    \label{eq:BUG_Einstein}
\end{align}
where%
\footnote{
The matter frame stress-energy tensor is found to be 
\begin{equation}
    \hat T_{\mu\nu} = \big (\hat X - \lambda_2\big)\hat g_{\mu\nu} + \partial_\mu \phi \partial_\nu \phi.
\end{equation}
}
\begin{align}
    T_{\mu\nu} = 
    \left( \frac{X}{\nu} - \nu \lambda_2\right) g_{\mu\nu}
    +
    \left( \frac{1-BX}{\nu^3} + \frac{B\lambda_2 }{\nu}  \right)
    \partial_\mu\phi\partial_\nu\phi .
    \label{eq:BUG_SET}
\end{align}
Interestingly, this effective stress tensor has the same algebraic scalar-fluid form encountered in related mimetic gravity constructions~\cite{jirousek2022mimetickessence}. Though the resemblance is only structural, where the physical distinction lies in the constraint determining $X(x,\phi)$ and in the behaviour of the Lagrange multipliers.

\subsubsection{BUG Vacuum Contributions}
It remains to determine the nature of the Lagrange multipliers $\lambda_{1,2}$. In ordinary UG, one identifies the Lagrange multiplier with a cosmological constant by taking the divergence of the Einstein equations (see Sec.~\ref{sec:UG}). The same calculation in the present bimetric case, however, yields a different result. Using the biscalar equation of motion \eqref{eq:BUG_ScalarEoM}, re-expressed in the Einstein frame, one finds that the BUG biscalar stress tensor \eqref{eq:BUG_SET} is not separately conserved,
\begin{equation}
    \nabla_\mu T^{\mu\nu}
    =
    -\nu \nabla^\nu \lambda_2 .
    \label{eq:BUG_T_noncons}
\end{equation}
This comes with no surprise, since the fixed-density construction is not fully diffeomorphism invariant \cite{wald:1984}. Taking the divergence of \eqref{eq:BUG_Einstein} and using the Bianchi identity therefore yields
\begin{equation}
\nabla_\mu\lambda_1
=
-\nu \nabla_\mu\lambda_2 .
\label{eq:BUG_balancelaw}
\end{equation}
Thus neither BUG multiplier need be constant. Their spacetime dependence is instead tied together by the balance law \eqref{eq:BUG_balancelaw}. Consequently, the final term in the scalar equation \eqref{eq:BUG_ScalarEoM} survives.

The balance law \eqref{eq:BUG_balancelaw} also carries an integrability condition. Since $\lambda_1$ is a scalar, the one-form $-\nu d\lambda_2$ must be exact \cite{Nakahara2003}. Taking an exterior derivative of $d\lambda_1=-\nu d\lambda_2$ gives $d\nu\wedge d\lambda_2=0$, or equivalently in components,
\begin{equation}
    \nabla_{[\mu}\nu\,\nabla_{\nu]}\lambda_2=0 .
\end{equation}
Thus, for a prescribed relative volume element $\nu(x)$, the second multiplier cannot be chosen as an arbitrary independent spacetime function. Rather, locally its gradient must be aligned with that of $\nu$. This condition is of course trivially satisfied on homogeneous FLRW backgrounds, where all quantities depend only on time.

The key novel aspect of generic BUG, therefore, is that it does not possess a conserved bimodular cosmological constant. In the special case where the relative volume element is constant, $\nu(x)=\nu$, the balance law becomes
\begin{equation}
    \nabla_\mu\left(\lambda_1+\nu\lambda_2\right)=0.
    \label{eq:Bi-UG_lambda_constraint}
\end{equation}
Only in this special constant-$\nu$ case may the bracketed combination be identified with a genuine bimodular cosmological constant, albeit one built from dynamical constituents. For a general fixed scalar density ratio, however,
\begin{equation}
    \Pi(x)
    \coloneqq
    \lambda_1(x)+\nu(x)\lambda_2(x)
\end{equation}
obeys
\begin{equation}
    \nabla_\mu \Pi
    =
    \lambda_2\nabla_\mu\nu .
    \label{eq:varying_vacuum}
\end{equation}
It is therefore better interpreted as an effective vacuum contribution whose spacetime variation is sourced by the prescribed relative volume element.

Motivated by these properties, we shall henceforth refer to any classical realisation of the construction whose dynamics prescribe the relative volume element, constrain the kinetic term to the form $X=X(x;\phi)$, and generate an effective vacuum contribution which need not be conserved as \textit{Bimodular Gravity}.\footnote{
The definition is deliberately dynamical, rather than tied to a particular off-shell presentation, because he same structure may be implemented by fixed density constraints, auxiliary vector densities, or equivalent covariant reformulations (see Sec.~\ref{sec:Covariant_Reformulation}).
}

\subsection{Bimetric HT Gravity Theory}
\label{subsec:BiHT}
We now turn to the HT implementation of the same bimetric theory. In the single-metric case, this formulation is classically equivalent to fixed-determinant unimodular gravity. In the present disformal bimetric setting, however, the two implementations do not lead to the same classical dynamics. We show that the resulting theory is therefore closer to a standard k-essence model, supplemented by a cosmological constant and a constant scalar-sector offset, whose perturbative stability can be classified in the usual way.

\subsubsection{HT Bimetric Constraints}
To preserve full diffeomorphism invariance in bimodular gravity, we instead implement HT constraints for both metrics, thereby defining the \textit{Bimetric Henneaux and Teitelboim} (BHT) theory, whose constraint sector in the bimodular action \eqref{eq:bimodular_gravity_action_general} reads
\begin{align}
    S_{BHT} \supset - \int d^4x \Bigg[&
    \lambda_1 \left( \sqrt{-g} - \partial_\mu\tau_1^\mu \right) 
    + \lambda_2 \left( \sqrt{-\hat g} - \partial_\mu\tau_2^\mu \right)
    \Bigg].
    \label{eq:eq:BHT_action}
\end{align}
Variation with respect to the auxiliary densities enforces the spacetime constancy of the multipliers,
\begin{align}
    \partial_\mu\lambda_1 = 0,
    \qquad 
    \partial_\mu\lambda_2 = 0,
    \label{eq:BHT_constant_lambdas}
\end{align}
so that $\lambda_{1,2}$ are individually integration constants, in contrast to BUG in which the multipliers obey a balance law \eqref{eq:BUG_balancelaw}. This is the first distinction from BUG; the second is kinematic. Variation with respect to $\lambda_{1,2}$ imposes the corresponding bimodular constraints
\begin{equation}
    \sqrt{-g}=\partial_\mu\tau_1^\mu,
    \qquad 
    \sqrt{-\hat g}=\partial_\mu\tau_2^\mu.
    \label{eq:BHT_constaints}
\end{equation}
Consequently, the relative volume element becomes
\begin{equation}
    \nu = \frac{\partial_\mu\tau_2^\mu}{\partial_\mu\tau_1^\mu} 
    = \sqrt{1-2B X},
    \label{eq:BHT_V}
\end{equation}
which, unlike BUG, remains fully dynamical. Consequently, the biscalar may keep its full dynamical freedom.

\subsubsection{The Dynamics of BHT}
\label{subsec:BHT_dynamics}
Variation of the BHT action \eqref{eq:eq:BHT_action} with respect to the biscalar yields
\begin{align}
    \left[B (\hat X  + \lambda_2) + 1 \right] \hat \Box \phi
    =
    - B\hat \nabla_\mu \hat X \hat \nabla^\mu \phi .
    \label{eq:BiHT_ScalarEoM}
\end{align}
Here we have used the defining BHT property that the Lagrange multipliers are integration constants \eqref{eq:BHT_constant_lambdas}. Consequently, the term proportional to $\hat\nabla_\mu\lambda_2$ vanishes identically.

Variation with respect to the Einstein-frame metric gives
\begin{align}
    G_{\mu\nu}+\lambda_1 g_{\mu\nu}
    =
    T_{\mu\nu},
    \label{eq:BHT_Einstein}
\end{align}
with
\begin{align}
    T_{\mu\nu} = 
    \left( \frac{X}{\nu}  - \nu \lambda_2\right) g_{\mu\nu}
    + \left( \frac{1-BX}{\nu^3} + \frac{B\lambda_2 }{\nu}  \right)
    \partial_\mu\phi\partial_\nu\phi .
    \label{eq:BHT_SET}
\end{align}
This is algebraically identical to the BUG stress tensor, but its interpretation is different. In BHT, $\lambda_1$ and $\lambda_2$ are separately constant, and the action is fully diffeomorphism invariant. Taking the divergence of \eqref{eq:BHT_Einstein} therefore gives
\begin{equation}
    \nabla_\mu T^{\mu\nu}=0 ,
    \label{eq:BHT_T_cons}
\end{equation}
so the Bianchi identity is satisfied in the standard way \cite{Hallam_2024}.

\subsubsection{BHT Vacuum Contributions}
The two constant vacuum contributions are thus separated in BHT. The first multiplier appears directly in the Einstein equations and may be identified with the cosmological constant,
\begin{equation}
    \Lambda \coloneqq \lambda_1 .
    \label{eq:BHT_lambda1}
\end{equation}
The second multiplier acts as a constant scalar potential in the matter frame,
\begin{equation}
    V_0\coloneq\lambda_2.
    \label{eq:BHT_lambda2}
\end{equation}
Hence BHT behaves as a unimodular gravity plus a disformally induced k-essence scalar, with the cosmological constant and scalar potential offset arising as distinct integration constants. This structure is close in spirit to mimetic k-essence \cite{jirousek2022mimetickessence} and generalized unimodular k-essence \cite{Barvinsky_2021} constructions.

Evidently, BHT is a classically distinct theory to BUG. In BUG, the scalar field is non-propagating, and the two multipliers are tied by the balance law \eqref{eq:BUG_balancelaw}, such that for a general fixed relative volume element no conserved combination $\Pi(x)$ exists. In BHT, by contrast, the two constants remain independently conserved, whilst the scalar retains its genuine dynamical degree of freedom. Thus, BHT does not fall within the class of Bimodular Gravity. It is a more standard k-essence with a standard k-essence speed of sound which we now classify.

\subsubsection{BHT Speed of Sound}
\label{subsubsec:BHT_sound_speed}
For the propagating scalar field of BHT, the scalar sound speed is obtained via 
\begin{equation}
    c_s^2 =\frac{K_X}{K_X + 2X K_{XX}},
    \label{eq:sound_speed_general}
\end{equation}
where subscripts denote partial derivatives with respect to $X$ \cite{Garriga_1999}. For BHT, the scalar kinetic function is read off as
\begin{align}
    K(\phi, X;\lambda_2) = \frac{X}{\sqrt{1-2BX}} -\sqrt{1-2BX}  \, V_0.
    \label{eq:BMG_kessence}
\end{align}
Evaluating the partial derivatives with respect to $X$ at fixed $\phi$ and $ \lambda_2$, one finds
\begin{align}
    K_X &= \frac{1-BX}{\nu^3} + \frac{B}{\nu}V_0,
    \label{eq:KX_HT} \\
    K_{XX} &= \frac{B (2-BX)}{\nu^5} + \frac{B^2}{\nu^3}V_0.
    \label{eq:KXX_HT}
\end{align}
Substituting \eqref{eq:KX_HT}-\eqref{eq:KXX_HT} into \eqref{eq:sound_speed_general} yields
\begin{equation}
    c_s^2=
    \nu^2\frac{ N}
    {D}, 
\label{eq:cs_explicit}
\end{equation}
where we have defined
\begin{equation}
    N \coloneq 1-BX +\nu^2  BV_0,
    \qquad
    D  \coloneq 1 + B X + \nu^2BV_0.
\end{equation}

The absence of gradient instabilities $c_s^2>0$ can be satisfied on two algebraic branches. On the positive-denominator branch, $D>0$, the condition requires
\begin{equation}
    V_0>
    \frac{BX-1}{B\nu^2}.
    \label{eq:BHT_grad_cond}
\end{equation}
There is also a negative-denominator branch, which requires
\begin{equation}
    V_0<-\frac{1+BX}{B\nu^2},
    \label{eq:BHT_neg_grad_cond}
\end{equation}
whereby gradient stability would also require the numerator to be negative. For $BX>0$, this follows immediately from \eqref{eq:BHT_neg_grad_cond}, since then
\begin{equation}
    (1-BX)+B\nu^2 V_0<-2BX<0 .
\end{equation}
For $BX<0$, however, the negativity of the numerator is not automatic and must be imposed separately. In any case, this branch has negative kinetic coefficient and is therefore discarded as ghost-like. We therefore restrict to the branch satisfying both $D>0$ and $N>0$, with $BX<1/2$, such that the biscalar is free of ghosts and gradient instabilities.

On this healthy branch, the sound speed can then be classified according to the value of $BX$ as
\begin{align}
    &\text{(i) } BX\to 1/2^-:
    && \nu^2\to0,
    && c_s^2\to0,
    && \text{dust-like limit},
    \nonumber\\[0.3em]
    &\text{(ii) } 0<BX<1/2:
    && 0<\nu^2<1,
    && 0<c_s^2<1,
    && \text{subluminal},
    \nonumber\\[0.3em]
    &\text{(iii) } BX=0:
    && \nu^2=1,
    && c_s^2=1,
    && \text{luminal},
    \nonumber \\[0.3em]
    &\text{(iv) } BX<0:
    && \nu^2>1,
    && c_s^2>1,
    && \text{superluminal}.
    \nonumber
    \label{eq:BHT_sound_speed_classes}
\end{align}
Note that superluminality \cite{Babichev_2008, Sawicki_2025} occurs when $BX<0$, which for $B>0$, requires $X<0$, corresponding to a spacelike kinetic term.

\subsection{Bimetric Fully Diffeomorphism-Invariant Unimodular Gravity Theory}
\label{subsec:BiDUG}
Alternatively, still under the wish to preserve full diffeomorphism invariance in the bimodular framework, one may instead employ the DUG formalism as to define \textit{Bimetric Diffeomorphism-invariant Unimodular Gravity} (BDUG), whose constraint sector becomes
\begin{equation}
    S_{BDUG} \supset 
    - \int d^4x 
    \Bigg[
    \sqrt{-g} \left( \lambda_1 + V_1^\mu\nabla_\mu \lambda_1 \right) 
    + \sqrt{-\hat g}\left( \lambda_2 + V_2^\mu \hat \nabla_\mu \lambda_2 \right) 
    \Bigg],
    \label{eq:BDUG_action}
\end{equation}
where $\nabla_{\mu}$ and $\hat\nabla_{\mu}$ denote the covariant derivatives compatible with $g_{\mu\nu}$ and $\hat g_{\mu\nu}$, respectively. Variation with respect to the auxiliary vectors enforces spacetime constancy of the multipliers,
\begin{equation}
    \nabla_\mu \lambda_1 = 0,
    \qquad 
    \hat \nabla_\mu\lambda_2 = 0,
    \label{eq:BHT_int_consts}
\end{equation}
so that, just as in BHT, $\lambda_{1,2}$ play the role of strict integration constants. Varying with respect to $\lambda_{1,2}$ yields the divergence conditions
\begin{equation}
    \nabla_\mu V^\mu_1 = 1,
    \qquad 
    \hat \nabla_\mu V^\mu_2 = 1,
\end{equation}
which entirely replace the fixed–determinant constraints by normalisation conditions on the auxiliary vectors. As a consequence, the relative volume element remains fully dynamical.

The shared imposition of constant $\lambda_{1,2}$ and fully dynamical $BX$ means that the BHT and BDUG theories are equivalent on shell. Hence, for all intents and purposes of this work, we treat them as identical and focus on the BHT formulation for notational convenience, with its classical dynamics taken to represent those of BDUG---neither of which classifies as a Bimodular Gravity theory.

\section{Cosmology}
\label{sec:cosmo}
Having established the distinct dynamics of BUG and BHT, we now investigate their cosmological implications. We impose the spatially flat FLRW ansatz in the bimetric setting, derive the corresponding background equations, and examine how each implementation shapes the effective expansion history. In particular, we reconstruct the dark energy equation of state and determine whether a crossing of the phantom divide may occur in either case.

\subsection{Bimetric FLRW Ansatz}
We specialise to a spatially flat, homogeneous, and isotropic cosmology for both geometries, displaying explicit FLRW forms for the Einstein metric and its disformally related counterpart~\cite{Moffat_2003}. In the Einstein frame,
\begin{equation}
    ds^2 = -N(t)^2dt^2 + a(t)^2 (dx^2 + dy^2 + dz^2),
\end{equation}
and the biscalar is homogeneous, $\phi = \phi(t)$, so $\partial_{\mu}\phi = (\dot{\phi}, 0, 0, 0)$ (where an overdot $\dot{}$ denotes $d/dt$). Using the disformal relation \eqref{eq:disformal_relation}, the components of the matter-frame metric are 
\begin{equation}
    \hat g_{00} = -N(t)^2 + B(\phi)\dot\phi^2,
    \qquad
    \hat g_{ij} = a(t)^2\delta_{ij}.
\end{equation}
Hence, only the time-time component is altered between the frames, whilst the scale factor remains the same. The matter-frame line element thus takes the form
\begin{align}
    d\hat s^2
    &= -\hat{N}(t)^2dt^2 + \hat{a}(t)^2(dx^2 + dy^2 + dz^2),
\end{align}
where 
\begin{equation}
    \hat{N}(t)^2 \coloneqq N(t)^2 - B(\phi)\dot{\phi}^2, \quad \hat{a}(t) \equiv a(t).
    \label{eq:disformal_lapse}
\end{equation}
Reality of the lapse requires $\hat N^2>0$, i.e. $N^2>B\dot\phi^2$, corresponding to the healthy domain $1-2BX>0$. For completeness, the relative volume element is then $\nu = \hat N/N$.

The proper times in each frame correspond to
\begin{equation}
    d\tau \coloneqq Ndt
    \qquad
    d\hat\tau \coloneqq \hat Ndt
    \label{eq:propertimes}
\end{equation}
We thus define the Einstein- and matter-frame Hubble rates as
\begin{equation}
    H \coloneqq \frac{1}{a}\frac{da}{d\tau} = \frac{1}{a}\frac{da}{Ndt},
    \qquad
    \hat H \coloneqq \frac{1}{a}\frac{da}{d\hat\tau} = \frac{1}{a}\frac{da}{\hat N dt},
\end{equation}
where, since $d\hat\tau = \nu  d\tau$, the two are related by
\begin{equation}
    H = \nu\hat H.
    \label{eq:H_Hhat_relation}
\end{equation}

\subsection{Cosmological Field Equations}
Reducing the bimetric action with general constraints \eqref{eq:bimodular_gravity_action_general} to the spatially flat FLRW ansatz yields 
\begin{align}
    S = \int dt \Bigg[& 
    -3 \frac{a \dot a^2}{N}  + a^3 \frac12 \frac{\dot{\phi}^2}{\hat N}  
    - \lambda_1 \left( N a^3 - \mathcal C_1 \right) - \lambda_2 \left( \hat N a^3 - \mathcal C_2 \right) 
    \Bigg].
    \label{eq:BiM_action_MSS}
\end{align}
Variation with respect to $\lambda_{1,2}$ reproduces the constraints
\begin{equation}
    N a^3= \mathcal C_1, 
    \qquad 
    \hat N a^3 = \mathcal C_2,
\end{equation}
whose specific forms depend on the chosen constraints. Variation with respect to $N$ gives the modified Friedmann equation
\begin{equation}
    3H^2 = \lambda_1 + \dfrac{1}{\nu} \left( \hat{X} +  \lambda_2 \right).
    \label{eq:ModFriedmann}
\end{equation}
Varying with respect to $a$ and eliminating $\lambda_1$ using \eqref{eq:ModFriedmann} gives the matter-frame Raychaudhuri equation 
\begin{equation}
    \frac{d\hat H}{d\hat\tau } = \frac{B \hat H}{\nu^2} \frac{dX}{d\hat\tau}
    +\frac{\hat X}{\nu}
    \left[
    B(\hat X - \lambda_2)
    - \frac{1}{\nu^2}
    \right],
    \label{eq:ray}
\end{equation}
whilst variation with respect to $\phi$ gives the background biscalar equation,
\begin{equation}
    \frac{d}{dt}\left[
    a^3 \frac{\dot\phi}{\hat N} \left(1 + B(\hat X  + \lambda_2)\right)
    \right]
    = 0 ,
    \label{eq:homo_biscalar_eom}
\end{equation}
which integrates once to give
\begin{equation}
    a^3\phi'
    \left(1 + B(\hat X + \lambda_2)\right)
    = C,
    \label{eq:homo_biscalar_eom_int}
\end{equation}
where $C$ is a scalar-charge integration constant. 
Equations~\eqref{eq:ray}-\eqref{eq:homo_biscalar_eom_int} together determine the background cosmology once the specific constraints are specified. We now look at the cosmologies of each of the theories in turn.

\subsection{The Cosmological Dynamics of BUG}
\label{subsec:BUG_cosmology}
Specialising BUG to the spatially flat FLRW background, we find that the background dynamics reduce to a single first-order equation which we then use to investigate the dark energy equation of state and the possibility of phantom-divide crossing in the Einstein and matter frames. Hereafter we denote differentiation with respect to proper time by a prime $'$.

\subsubsection{Homogeneous BUG Constraints}
The fixed-density constraints \eqref{eq:BUG_constaints} read
\begin{equation}
    N a^3 = \varepsilon_1,
    \qquad
    \hat N a^3 = \varepsilon_2,
\end{equation}
so that the relative volume element is simply the ratio of the two lapse functions,
\begin{equation}
    \nu(\hat\tau)
    =
    \frac{\hat N}{N}
    =
    \frac{\varepsilon_2(\hat\tau)}{\varepsilon_1(\hat\tau)},
    \label{eq:BUG_nu_FLRW}
\end{equation}
which themselves are prescribed via the fixed densities $\varepsilon_{1,2}(\hat\tau)$. This is the defining feature of the BUG background \eqref{eq:BX-const}: $\nu$ is not a dynamical variable but an externally prescribed function, and the entire scalar sector is determined by it. The kinematic constraints \eqref{eq:kin_lock}-\eqref{eq:kin_lock_deriv} fix the biscalar velocity and acceleration algebraically,
\begin{equation}
    \phi'
    =
    \sigma\sqrt{\frac{1-\nu^2}{B\nu^2}},
    \qquad
    \hat X
    =
    \frac{1}{2}\phi'^2
    =
    \frac{1-\nu^2}{2B\nu^2},
    \qquad
    X' = -\frac{\nu}{B}\nu',
    \label{eq:BUG_phiprime_hatX}
\end{equation}
where $\sigma = \pm 1$ labels the scalar branch. The evolution of the biscalar is inherited entirely from the prescribed $\nu(\hat\tau)$.

\subsubsection{BUG Background Expansion} 
With $\phi'$ and $\hat X$ already fixed, the homogeneous biscalar equation \eqref{eq:homo_biscalar_eom_int} no longer propagates the field but instead determines the second multiplier,
\begin{equation}
    \lambda_2
    =
    \frac{C}{B a^3\phi'}
    - \frac1B
    - \hat X,
\label{eq:BUG_lambda2_reconstruction_raw}
\end{equation}
which, using \eqref{eq:BUG_phiprime_hatX}, becomes
\begin{equation}
    \lambda_2
    =
    \mu \frac{\nu}{a^3\sqrt{1-\nu^2}}
    -
    \frac{1+\nu^2}{2B\nu^2},
    \qquad
    \mu \coloneqq \frac{\sigma C}{\sqrt{B}}.
    \label{eq:BUG_lambda2_reconstruction}
\end{equation}
The first multiplier then follows from the Friedmann equation, using
\eqref{eq:H_Hhat_relation},
\begin{equation}
    \lambda_1
    =
    3H^2 - \frac{\hat X + \lambda_2}{\nu}
    =
    3H^2
    -
    \frac{\mu}{a^3\sqrt{1-\nu^2}}
    +
    \frac{1}{B\nu}.
\label{eq:BUG_lambda1_reconstruction}
\end{equation}
Neither multiplier is therefore an independent---or even a combined---cosmological constant in the general BUG background. Both run with time, at rates set by $\mu$, the expansion rate, and the prescribed relative volume element.

Substituting \eqref{eq:BUG_phiprime_hatX} and \eqref{eq:BUG_lambda2_reconstruction} into the Raychaudhuri equation \eqref{eq:ray}, every purely algebraic term cancels and the dynamics collapses to a single first-order equation,
\begin{equation}
    \hat H'
    =
    -\frac{\nu'}{\nu}\hat H
    -
    \frac{\mu}{2a^3}\frac{\sqrt{1-\nu^2}}{\nu^2},
    \label{eq:BUG_Hhat_prime}
\end{equation}
which, together with
\begin{equation}
    a' = a\hat H,
    \label{eq:BUG_a_prime}
\end{equation}
closes the BUG background system.

In the Einstein frame it takes the simpler form
\begin{equation}
    H'
    =
    -\frac{\mu}{2a^3}\frac{\sqrt{1-\nu^2}}{\nu},
    \label{eq:BUG_H_prime}
\end{equation}
and, using $a' = aH/\nu$, may be recast directly as an equation in the scale factor,
\begin{equation}
    \frac{dH^2}{da}
    =
    -\mu \frac{\sqrt{1-\nu^2(a)}}{a^4}.
    \label{eq:BUG_H2_a}
\end{equation}
The background has thus been reduced to a single quadrature. Imposing $H(a=1) = H_0$ gives
\begin{equation}
    H^2(a)
    =
    H_0^2
    +
    \mu\int_a^1
    \frac{\sqrt{1-\nu^2(\bar a)}}{\bar a^4}
     d\bar a,
    \label{eq:BUG_H_solution}
\end{equation}
or, in dimensionless form with $E(a) \coloneqq H(a)/H_0$,
\begin{equation}
    E^2(a)
    =
    1
    +
    \Gamma\int_a^1
    \frac{\sqrt{1-\nu^2(\bar a)}}{\bar a^4}
     d\bar a,
    \qquad
    \Gamma \coloneqq \frac{\mu}{H_0^2}.
    \label{eq:BUG_E_solution}
\end{equation}
Once $\nu(a)$ is given the expansion history is fully determined, with the single parameter $\Gamma$ setting the amplitude and sign of the departure from de Sitter.

\subsubsection{Frame Dependence of Phantom Crossing in BUG}
\label{sec:BUG_EOS}
The frame in which observables are defined depends on which metric the matter fields couple to. If ordinary matter is coupled to the matter metric $\widehat g_{\mu\nu}$, $S_{\rm m}[\widehat g_{\mu\nu}, \psi]$, then the expansion history inferred by physical observers is the matter-frame one. Since the two FLRW metrics share the same scale factor but differ in their proper times, up to an irrelevant constant normalisation of the dimensionless Hubble rates, we have
\begin{equation}
    \widehat E^2(a)=\frac{E^2(a)}{\nu^2(a)} .
\end{equation}
The effective matter-frame equation of state reconstructed from the background expansion is therefore
\begin{equation}
    \widehat w_{\rm BUG}(a)
    \coloneqq
    -1-\frac{1}{3}\frac{d\ln \widehat E^2}{d\ln a}.
    \label{eq:wBUG_matter_def}
\end{equation}
Equivalently,
\begin{equation}
    \widehat w_{\rm BUG}(a)
    =
    -1-\frac{1}{3}\frac{d\ln E^2}{d\ln a}
    +\frac{1}{3}\frac{d\ln \nu^2}{d\ln a}
    =
    w_{\rm BUG}(a)
    +\frac{1}{3}\frac{d\ln \nu^2}{d\ln a}.
    \label{eq:wBUG_frame_relation}
\end{equation}

For the BUG solution \eqref{eq:BUG_E_solution}, one finds
\begin{align}
    \widehat w_{\rm BUG}(a)+1
    &=
    w_{\rm BUG}(a)+1
    +\frac{1}{3}\frac{d\ln \nu^2}{d\ln a},
    \label{eq:wBUG_hat_plus_one}
    \\[0.5em]
    w_{\rm BUG}(a)+1
    &=
    \frac{\Gamma}{3E^2(a)}
    \frac{\sqrt{1-\nu^2(a)}}{a^3}.
    \label{eq:wBUG_einstein_plus_one}
\end{align}
For any physical BUG background, $E^2>0$ and $0<\nu^2\leq 1$. Hence $\sqrt{1-\nu^2}/a^3$ is non-negative, and the sign of $w_{\rm BUG}+1$ is fixed entirely by the sign of $\Gamma$. The Einstein-frame BUG expansion may therefore be quintessence-like for $\Gamma>0$ or phantom-like for $\Gamma<0$, but it cannot dynamically cross the phantom divide. 

The matter-frame equation of state, however, is different. Its distance from the phantom divide is
\begin{equation}
    \widehat w_{\rm BUG}(a)+1
    =
    \frac{1}{3}
    \left[
    \frac{\Gamma}{E^2(a)}
    \frac{\sqrt{1-\nu^2(a)}}{a^3}
    +
    \frac{d\ln \nu^2}{d\ln a}
    \right].
    \label{eq:wBUG_hat_crossing_condition}
\end{equation}
Consequently, a matter-frame phantom crossing occurs whenever
\begin{equation}
    \frac{d\ln \nu^2}{d\ln a}
    =
    -\frac{\Gamma}{E^2(a)}
    \frac{\sqrt{1-\nu^2(a)}}{a^3}.
    \label{eq:wBUG_hat_crossing}
\end{equation}
Thus the prescribed evolution of the relative volume element can drive
$\widehat w_{\rm BUG}+1$ through zero even though $w_{\rm BUG}+1$ has a fixed sign
in the Einstein frame.

This crossing is therefore not an intrinsic Einstein-frame phantom crossing. Rather, it is a frame-dependent effect induced by the changing relative volume element between the Einstein and matter metrics. If matter is universally coupled to $\widehat g_{\mu\nu}$, then observers would reconstruct the matter-frame quantity $\widehat w_{\rm BUG}$ and could infer a genuine background-level crossing of the phantom divide. Its origin, however, would lie in the time-dependent relative volume element between the two frames, not in a sign change of the Einstein-frame BUG source.

\subsubsection{Power-Law Relative Volume Element}
\label{sec:BUG_powerlaw_ansatz}
The residual freedom in BUG is carried entirely by the prescribed relative volume element
$\nu(a)$. To illustrate the frame dependence of the reconstructed equation of state, we
consider the simple power-law profile
\begin{equation}
    \nu^2(a)=1-q a^n,
    \qquad
    0<q a^n<1,
    \label{eq:BUG_powerlaw_nu}
\end{equation}
where the inequality specifies the physical domain in which the matter-frame lapse remains real. For $0<q<1$ and $n>0$, this condition is automatically satisfied over the interval $0<a\leq 1$. For $n<0$, the ansatz should instead be regarded as a late-time profile over the range in which $q a^n<1$.

With \eqref{eq:BUG_powerlaw_nu}, the solution \eqref{eq:BUG_E_solution} reads\footnote{For the special case $n=6$, the corresponding solution is instead
    \begin{equation}
        E^2(a)=1+\Gamma\sqrt q \ln\frac{1}{a}.
        \label{eq:BUG_powerlaw_solution_n6}
    \end{equation}
}
\begin{equation}
    E^2(a)
    =
    1
    +
    \Gamma\sqrt q 
    \frac{1-a^{n/2-3}}{n/2-3},
    \qquad
    n\neq 6.
    \label{eq:BUG_powerlaw_solution}
\end{equation}
The Einstein-frame equation of state therefore obeys
\begin{equation}
    w_{\rm BUG}(a)+1
    =
    \frac{\Gamma\sqrt q}{3E^2(a)}
    a^{n/2-3}.
    \label{eq:wBUG_powerlaw_Einstein}
\end{equation}
The matter-frame correction is
\begin{equation}
    \frac{1}{3}\frac{d\ln\nu^2}{d\ln a}
    =
    -\frac{nq a^n}{3(1-q a^n)}.
    \label{eq:BUG_powerlaw_frame_term}
\end{equation}
Hence the matter-frame equation of state is
\begin{equation}
    \widehat w_{\rm BUG}(a)+1
    =
    \frac{1}{3}
    \left[
    \frac{\Gamma\sqrt q}{E^2(a)}
    a^{n/2-3}
    -
    \frac{nq a^n}{1-q a^n}
    \right].
    \label{eq:wBUG_powerlaw_matter}
\end{equation}
A matter-frame phantom crossing occurs at any scale factor $a=a_\star$ satisfying
\begin{equation}
    \frac{\Gamma\sqrt q}{E^2(a_\star)}
    a_\star^{n/2-3}
    =
    \frac{nq a_\star^n}{1-q a_\star^n}.
    \label{eq:BUG_powerlaw_crossing_condition}
\end{equation}
Since $E^2>0$ and $1-q a^n>0$ in the physical domain, this condition can be satisfied only when $\Gamma$ and $n$ have the same sign, $\Gamma n>0$.

The direction of the crossing is then controlled by the sign of $n$. For $n>0$ and $\Gamma>0$, the Einstein-frame solution is quintessence-like, whilst the matter-frame correction is negative. The matter frame may therefore cross from quintessence-like to phantom-like behaviour if the evolution of $\nu$ becomes sufficiently important. Conversely, for $n<0$ and $\Gamma<0$, the Einstein-frame solution is phantom-like, whilst the matter-frame correction is positive. This is the branch on which a matter-frame observer can reconstruct a transition from phantom-like to quintessence-like expansion.

\begin{figure}[h!]
    \centering
    \includegraphics[width=0.95\linewidth]{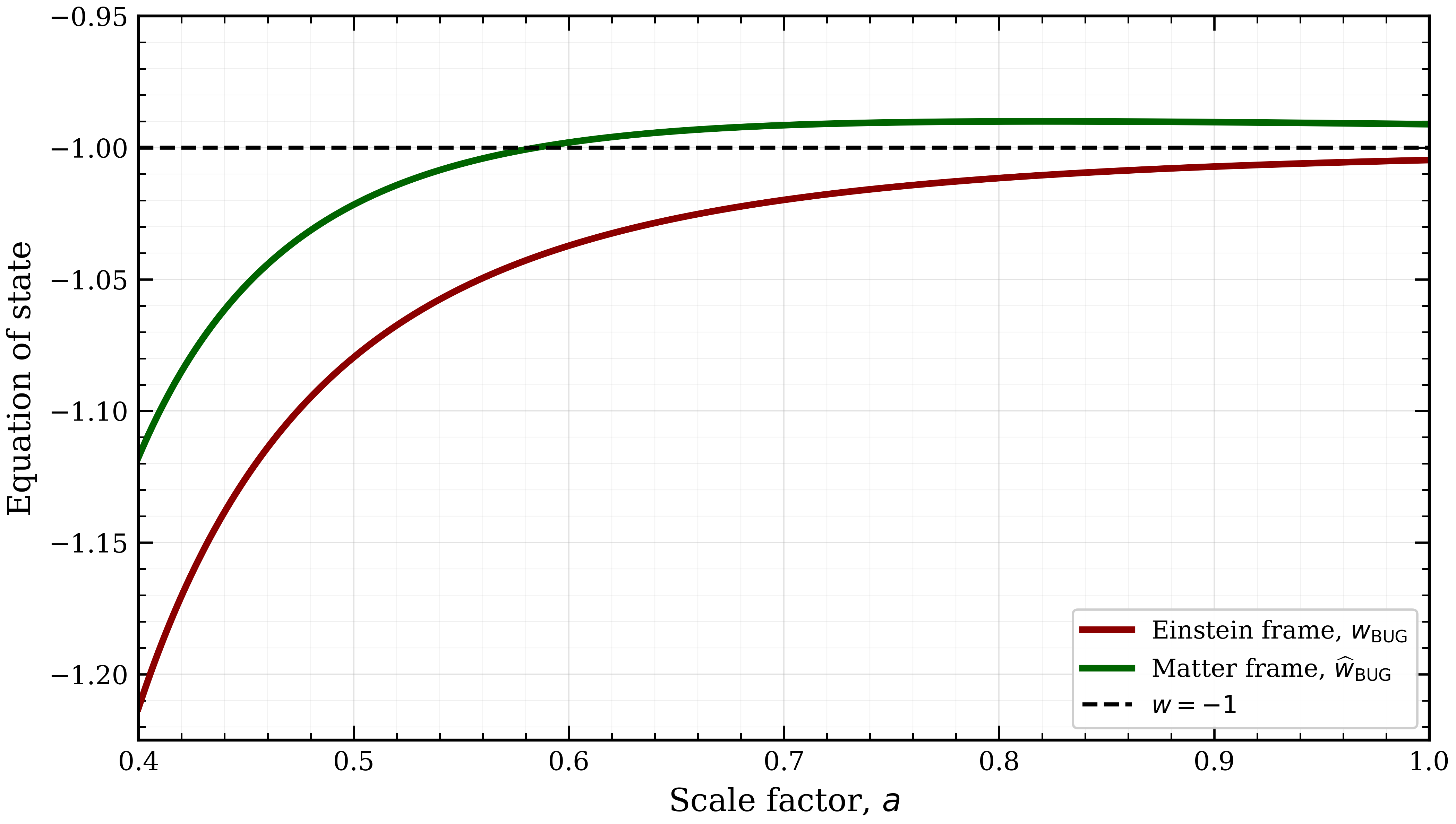}
    \caption{Frame dependence of the reconstructed BUG equation of state for the power-law relative volume element $\nu^2(a)=1-q a^n$. The two curves correspond to the same underlying BUG background, with representative parameter choice $q=0.02$, $n=-2$, and $\Gamma=-0.10$. In the Einstein frame, $w_{\rm BUG}$ remains below the phantom divide throughout the interval shown, as its sign is fixed by $\Gamma<0$. In the matter frame, however, the additional matter-frame contribution drives the reconstructed equation of state $\widehat w_{\rm BUG}$ through $-1$, giving a phantom-to-quintessence transition at $a_\star\simeq0.585$ for this illustrative parameter choice. The location of the crossing is, of course, parameter-dependent; the point is that the crossing is present only in the matter-frame reconstruction.}
    \label{fig:placeholder}
\end{figure}

To illustrate this frame-dependent crossing explicitly, we plot in Fig.~\ref{fig:placeholder} the Einstein- and matter-frame equations of state for the same BUG background solution. We choose the inverse-power branch $n<0$ with $\Gamma<0$, for which the Einstein-frame source is everywhere phantom-like. As expected from \eqref{eq:wBUG_powerlaw_matter}, the Einstein-frame equation of state remains below $-1$ throughout the interval shown and therefore does not cross the phantom divide. By contrast, the matter-frame reconstruction does cross from phantom-like to quintessence-like behaviour. This crossing is not caused by a sign change of the Einstein-frame BUG source, but by the additional matter-frame contribution induced by the evolving relative volume element. Thus the same physical background can appear crossing or non-crossing depending on the frame in which the equation of state is reconstructed.

\newpage
\subsection{The Cosmological Dynamics of BHT}
\label{sec:BHT_cosmology}
We now turn to the BHT realisation of bimodular gravity. Unlike BUG, the HT constraints do not prescribe the relative volume element, but instead enforce the two bimodular multipliers to be independent constants. The biscalar therefore remains fully dynamical, and the resulting theory is a more standard shift-symmetric k-essence model supplemented by a cosmological constant and a constant scalar-sector offset. We shall see that, whilst this branch can give non-trivial k-essence cosmology, it cannot dynamically cross the phantom divide.

\subsubsection{Homogeneous BHT Constraints}
Under the FLRW ansatz, with $\tau_i^\mu=\tau_i^\mu(t)$, the volume constraints become
\begin{equation}
    Na^3=\dot\tau_1,
    \qquad
    \widehat N a^3=\dot\tau_2.
    \label{eq:BHT_flrw_conts_t}
\end{equation}
Unlike in BUG, the auxiliary densities $\tau_i$ are dynamical objects. Their derivatives are not prescribed functions, and hence the relative volume element is not fixed a priori. The scalar therefore remains a genuine dynamical degree of freedom. The defining BHT condition is instead the constancy of the two multipliers,
\begin{equation}
    \lambda_1=\Lambda,
    \qquad
    \lambda_2=V_0,
    \label{eq:BHT_constants_cosmology}
\end{equation}
where $\Lambda$ is the usual unimodular cosmological constant and $V_0$ is a constant offset in the scalar sector.

\subsubsection{BHT Background Expansion}
For completeness, the Friedmann equation gives the algebraic constraint
\begin{equation}
    3\nu^2\widehat H^2
    =
    \Lambda
    +
    \frac{1}{\nu}
    \left(
        \frac{1}{2}\phi'^2+V_0
    \right),
    \label{eq:BHT_friedmann_constraint}
\end{equation}
though this has already been incorporated within the general matter-frame Raychaudhuri equation, which becomes
\begin{equation}
    \widehat H'
    =
    B\nu^2\widehat H\phi'\phi''
    -
    \frac{\phi'^2}{2\nu}
    \left[
        1+BV_0+\frac{B}{2}\phi'^2
    \right].
    \label{eq:BHT_Hhat_ODE_phi_second}
\end{equation}
The biscalar equation follows by differentiating the first integral \eqref{eq:homo_biscalar_eom_int}. Since $V_0$ is constant, this gives
\begin{equation}
    \phi''
    =
    -3\widehat H\phi'
    \frac{
        1+B\left(V_0+\frac{1}{2}\phi'^2\right)
    }{
        1+B\left(V_0+\frac{3}{2}\phi'^2\right)
    }.
    \label{eq:BHT_phi_ODE}
\end{equation}
Together with
\begin{equation}
    a'=a\widehat H,
    \label{eq:BHT_a_ODE}
\end{equation}
equations \eqref{eq:BHT_Hhat_ODE_phi_second}-\eqref{eq:BHT_a_ODE} define the BHT background system.

\subsubsection{Absence of Phantom Crossing in BHT}
The possibility of phantom crossing in BHT may be determined directly from the matter-frame Raychaudhuri equation. In the FLRW background, we find that the numerator and denominator of the scalar speed become 
\begin{equation}
    N = 1+B\left(V_0+\frac12\phi'^2\right),
    \qquad
    D = 1+B\left(V_0+\frac32\phi'^2\right).
\end{equation}
We then find that the scalar equation \eqref{eq:BHT_phi_ODE} takes the compact form
\begin{equation}
    \phi''=-3\widehat H\phi'\frac{N}{D}.
\end{equation}
Substituting this into \eqref{eq:BHT_Hhat_ODE_phi_second} gives
\begin{equation}
    \widehat H'
    =
    -\phi'^2 A
    \left[
        \frac{1}{2\nu}
        +
        \frac{3B\nu^2\widehat H^2}{D}
    \right].
    \label{eq:BHT_Hprime_sign}
\end{equation}
Thus, on the branch free of ghosts and gradient instabilities, we have $N,D>0$ (see Sec.~\ref{subsubsec:BHT_sound_speed}). Since $B>0$, $\nu>0$, and $\widehat H^2>0$, the bracket in \eqref{eq:BHT_Hprime_sign} is positive. Hence
\begin{equation}
    \widehat H'\leq 0,
\end{equation}
with equality only on the frozen-scalar branch $\phi'=0$ or at the boundary $N=0$.

The matter-frame dark energy equation of state satisfies
\begin{equation}
    \widehat w_{\rm BHT}+1
    =
    -\frac{2}{3}\frac{\widehat H'}{\widehat H^2}.
\end{equation}
Therefore, throughout the healthy rolling branch,
\begin{equation}
    \widehat w_{\rm BHT}+1\geq0.
\end{equation}
BHT can approach the de Sitter limit $\widehat w_{\rm BHT}=-1$ as $\phi'\to0$, but it cannot dynamically cross the phantom divide. Thus, unlike BUG, the BHT cosmology does not naturally realise a phantom crossing; it behaves as a more standard shift-symmetric k-essence theory.

\section{Diffeomorphism Invariant Bimodular Gravity}
\label{sec:Covariant_Reformulation}

So far we have made explicit the inequivalence between unimodularising bimetric gravity with the fixed-determinant constraints, BUG, and HT-type constraints, BHT. The former realises a genuinely novel bimodular theory, but one with broken diffeomorphism invariance. 
This prompts a natural question: can we construct a fully diffeomorphism-invariant completion that preserves the BUG balance law \eqref{eq:BUG_balancelaw} and its associated kinematic constraint on $X$ \eqref{eq:kin_lock}? In the remainder of this section we answer this question by developing a diffeomorphism-invariant formulation of bimodular gravity.

\subsection{Motivation}
For quantisation and for coupling to matter sectors in a standard covariant path integral, it is generally preferred to work with a formulation that is fully diffeomorphism invariant at the level of the action~\cite{2015-Bufalo}. In the single-metric case, one may pass from UG to HT (or DUG) without concern, for they are classically equivalent and reproduce the same field equations as $\mathrm{GR}\!+\!\Lambda$ \cite{1989-Henneaux, 2015-Bufalo, Jirousek_2023}. In the present disformal bimetric setting, however, this equivalence is broken. 

The general attempt to covariantise bimodular gravity by importing the HT/DUG formalism separately for each metric enforces both $\lambda_1$ and $\lambda_2$ to be strict integration constants. Conversely, in BUG, the individual multipliers remain dynamically correlated through the balance law. Treating the two HT sectors independently therefore overconstrains the multipliers---freezing both $\lambda_{1,2}$---and simultaneously underconstrains the relative volume element, leaving $X$ as a fully dynamical variable. The biscalar then behaves as a standard, freely propagating degree of freedom with a genuine second-order evolution equation.

Hence, two formalisms that are equivalent in the single-metric case become on-shell inequivalent once two disformally related metrics are present. This motivates the search for a diffeomorphism-invariant reformulation that is tailored specifically to the bimodular structure which:
\begin{enumerate}\setlength\itemsep{0.0002pt}
    \item realises full diffeomorphism invariance off shell;
    \item enforces the balance relation $\nabla_\mu \lambda_1 = -\nu\nabla_\mu\lambda_2$ on shell;
    \item reproduces the BUG kinematic constraint on $\nu$ and subsequently on $X=X(x;\phi)$;
    \item reduces to BUG in an appropriate gauge, and hence is classically equivalent to BUG.
\end{enumerate}
In what follows we present precisely such an action. One may regard it as a diffeomorphism-invariant completion of unimodularised bimetric scalar-tensor gravity, whereby full covariance is restored and the physical content of BUG is retained on shell.

\subsection{Restoring Diffeomorphism Invariance}
\label{subsec:proposed_action}

To this end, following the Stueckelberg covariantisation of generalised unimodular gravity~\cite{Barvinsky_2021} and the standard use of Stueckelberg fields to restore broken gauge symmetries \cite{Stueckelberg:1938zz}, we introduce four scalar Stueckelberg fields $Z^A(x)$, $A=0,\ldots,3$, and define the Jacobian density
\begin{equation}
    \mathcal J[Z]
    \coloneq
    \det\!\left(\partial_\mu Z^A\right).
\end{equation}
Let $\bar\epsilon_1(Z)$ and $\bar\epsilon_2(Z)$ be two fixed volume densities on the internal $Z^A$-space, with
\begin{equation}
    \bar\epsilon_2(Z)=\bar\nu(Z)\bar\epsilon_1(Z).
\end{equation}
Their pullbacks to spacetime define the scalar densities
\begin{equation}
    \mathcal E_i[Z]
    \coloneq
    \bar\epsilon_i(Z)\mathcal J[Z],
    \qquad i=1,2.
\end{equation}
Equivalently, $\mathcal E_i[Z]$ are the pullbacks of two fixed four-forms on the internal Stueckelberg space. The diffeomorphism-invariant BUG constraint sector is then
\begin{equation}
    S
    \supset
    -\int d^4x
    \left[
    \lambda_1\left(\sqrt{-g}-\mathcal E_1[Z]\right)
    +
    \lambda_2\left(\sqrt{-\hat g}-\mathcal E_2[Z]\right)
    \right].
    \label{eq:DBUG_action}
\end{equation}
This action is now fully diffeomorphism invariant.%
\footnote{
This essentially mirrors the Stueckelberg logic used in generalized unimodular gravity as k-essence \cite{Barvinsky_2021}, but not its dynamics. There the covariantised theory contains a genuine k-essence scalar; here the Stueckelberg fields restore diffeomorphism invariance whilst preserving the BUG constraint and multiplier balance law. In fact, BHT is closer to \cite{Barvinsky_2021}, since both contain a genuine k-essence-type scalar with standard perturbative sound speed. In BHT, however, the two HT multipliers remain separate constants, $\Lambda=\lambda_1$ and $V_0=\lambda_2$.
}

Variation with respect to the multipliers imposes
\begin{equation}
    \sqrt{-g}=\mathcal E_1[Z],
    \qquad
    \sqrt{-\hat g}=\mathcal E_2[Z].
\end{equation}
Taking the ratio gives
\begin{equation}
    \frac{\sqrt{-\hat g}}{\sqrt{-g}}
    =
    \frac{\mathcal E_2[Z]}{\mathcal E_1[Z]}
    =
    \nu(Z).
    \label{eq:stuck_rve}
\end{equation}
Thus the relative volume element is prescribed covariantly. In the unitary gauge
\begin{equation}
    Z^A=x^A,
\end{equation}
one has $\mathcal J=1$, $\mathcal E_i=\bar\epsilon_i(x)$, and hence
\begin{equation}
    \bar\nu(x)
    =
    \frac{\bar\epsilon_2(x)}{\bar\epsilon_1(x)}
    =
    \sqrt{1-2BX}.
\end{equation}
The usual BUG kinematic constraint is therefore recovered:
\begin{equation}
    X
    =
    \frac{1-\bar\nu^2(x)}{2B}.
\end{equation}

It remains to check the multiplier equation. Let $e^\mu{}_A$ denote the inverse Jacobian, satisfying $e^\mu{}_A \partial_\mu Z^B=\delta_A{}^B$. Since $\mathcal E_i[Z]$ are pullbacks of four-forms, their variations are total derivatives,
\begin{equation}
    \delta \mathcal E_i
    =
    \partial_\mu
    \left(
    \mathcal E_i e^\mu{}_A \delta Z^A
    \right).
\end{equation}
Variation of \eqref{eq:DBUG_action} with respect to $Z^A$ therefore gives
\begin{equation}
    \mathcal E_1 e^\mu{}_A\partial_\mu\lambda_1
    +
    \mathcal E_2 e^\mu{}_A\partial_\mu\lambda_2
    =0,
\end{equation}
which, for $\mathcal J \neq0$, is equivalent to
\begin{equation}
    \partial_\mu\lambda_1
    +
    \nu(Z)
    \partial_\mu\lambda_2
    =
    0,
\end{equation}
where we have used \eqref{eq:stuck_rve}. In the unitary gauge, $\nu(Z)=\bar \nu(x)$, and this becomes precisely the BUG balance law \eqref{eq:BUG_balancelaw}. Consequently, the effective vacuum contribution $\Pi(x)$ obeys \eqref{eq:varying_vacuum} and is therefore not conserved for a varying relative volume element.

This construction is therefore to be understood as a Stueckelberg completion of Bimodular Gravity. The Stueckelberg fields restore full diffeomorphism invariance, whilst their equations of motion reproduce the BUG multiplier balance law and restricted dynamics $X=X(x;\phi)$.

\section{Conclusion and Discussion}
\label{sec:conclusion}

We have unimodularised a simple bimetric scalar--tensor theory, in which gravity and matter couple to two disformally related metrics, by constraining the volume element of each. Although such unimodular formalisms dynamically coincide when only one metric is present, the procedure becomes inequivalent here. The fixed-determinant route prescribes the relative volume element $\nu(x)$, which in turn locks the scalar kinetic term to a function of spacetime, $X = X(x;\phi)$, and ties the two Lagrange multipliers $\lambda_{1,2}$ into a single vacuum contribution $\Pi(x)=\lambda_1(x) + \nu(x)\lambda_2(x)$ that runs as $\nabla_\mu\Pi(x) \propto\nabla_\mu\nu(x)$ rather than remaining a rigid constant. We have termed this class of dynamics Bimodular Gravity. A unique cosmological result of bimodularity is a frame-dependent crossing of the phantom divide: whilst impossible in the Einstein-frame, an observer coupled to the matter metric would see $w_{\rm DE}$ dynamically cross $-1$, driven entirely by the evolving ratio between the two volumes. The alternative route, which retains full diffeomorphism invariance through Henneaux--Teitelboim (HT) constraints, does not realise this behaviour. Instead, it returns a more standard k-essence theory carrying a genuine cosmological constant and a constant scalar offset in the matter frame, with no phantom crossing. Finally, we have given a Stueckelberg completion that restores diffeomorphism invariance to Bimodular Gravity whilst reproducing its dynamics on shell.

Thus, Bimodular Gravity supplies a simple geometric route to the recently inferred phantom crossing, following from the evolving ratio between the two metrics that matter and gravity respectively see. Furthermore, the unimodular reinterpretation of $\Lambda$ as an integration constant is carried one step further, with the constant being replaced by a contribution that evolves with the very volume element that drives the crossing, so that the two features share a common origin. Of course, $\nu(x)$ is prescribed rather than derived, so the crossing is engineered by a phenomenological choice rather than predicted; and, as with ordinary unimodular gravity, nothing here speaks to the smallness of the vacuum energy itself.

It should be emphasised that the bimodular constraint \eqref{eq:kin_lock} arises specifically from imposing two independent fixed-determinant constraints. If only one metric were constrained, i.e. if one unimodularised either $g_{\mu\nu}$ or $\hat g_{\mu\nu}$ alone, then the determinant of the remaining metric would be free to absorb variations in the relative volume element \eqref{eq:mathcalV}, and the biscalar kinetic term would not be prescribed. Similarly, if mixed constraints were implemented, with one fixed-determinant constraint and one HT constraint, the biscalar dynamics would remain free because the relative volume element would not be completely fixed. In that case, the constancy of the HT multiplier would, through the balance law \eqref{eq:BUG_balancelaw}, enforce the constancy of the remaining multiplier as well. The resulting theory would therefore fall on the side of a particular, but more standard, k-essence model, rather than Bimodular Gravity. Only when the two metric determinants are fixed, or equivalently fixed up to their covariant Stueckelberg completion, is the relative volume element prescribed and the bimodular kinetic constraint obtained.

It is also worth noting that the HT implementation admits a further, multiplier-free presentation. In \cite{Jirou_ek_2019}, unimodularity is obtained through a mimetic-like metric redefinition involving a vector field of conformal weight four, yielding a Weyl-invariant higher-derivative vector--tensor theory which reduces, in gauge-invariant variables, to the HT formulation. In the bimetric setting, applying such a construction would therefore amount to a Weyl-invariant rewriting of the BHT-type theory rather than of BUG. The two integration constants would remain independently conserved, whilst the biscalar would remain fully dynamical. 

\newpage

The dynamics of Bimodular Gravity sits close to, but distinct from, Mimetic Gravity~\cite{Chamseddine_2013, Chamseddine_2014} and Dust of Dark Energy \cite{Lim_2010}. In all three cases the scalar kinetic density is fixed algebraically, and hence the scalar sector is not described by an ordinary freely propagating k-essence mode. The similarities are nevertheless hierarchical. For constant $B$ and constant $\nu$, the bimodular constraint reduces to $X=\mathrm{const}$, mimicking the mimetic constant-norm constraint, up to conventions. Allowing $B=B(\phi)$, and more generally $\nu(\phi)$, instead gives $X=X(\phi)$, which is the kinematic structure underlying Dust of Dark Energy. Here, the pressure gradient is then aligned with the scalar flow, energy follows timelike geodesics, and the physical sound speed vanishes identically. Bimodular Gravity, however, corresponds to the further extension in which the prescribed relative volume element is a fixed spacetime density, $\nu(x)$. Even with general $B(\phi)$, the kinetic term is then locked to $X=X(x;\phi)$, so space-time gradients of $\nu$ source a non-zero acceleration of the scalar flow. Consequently, Bimodular Gravity is neither standard k-essence nor generically dusty dark energy. Indeed, its perturbations remain constrained rather than freely propagating, but the dust argument for an identically vanishing sound speed no longer applies.

It is likewise distinct from the mimetic k-essence construction \cite{jirousek2022mimetickessence}. There, the mimetic mixing of a scalar sector with a unimodular-type sector is shown to be on-shell equivalent to GR coupled to a k-essence scalar, whose overall scale appears as a conserved integration constant. Bimodular Gravity differs in both respects. Firstly, its scalar sector is not a freely propagating k-essence mode. Moreover, the vacuum contribution is inherited from two unimodular multipliers which, for generic $\nu(x)$, are not individually conserved and cannot be absorbed into a single constant---$\Pi(x)$ is instead a spacetime-dependent vacuum contribution.

Throughout, we have neglected the effects of matter, which must of course be included to take the cosmology seriously. The metric to which it couples is a genuine question, but the frame-dependent phantom crossing found here motivates a coupling to the matter metric $\hat g_{\mu\nu}$, with $S_{\rm m} = S_{\rm m}[\hat g_{\mu\nu},\psi]$. If matter then couples disformally to $\hat g_{\mu\nu}$, the viability of the construction would be further subject to local and screening constraints which we do not address here. A fuller assessment must therefore await a treatment of perturbations and ultimately a confrontation with data, to which we hope to return. It would also be desirable to render the prescribed $\nu(x)$ with a physical origin, so that the crossing is predicted rather than engineered. Finally, since part of the appeal of unimodular gravity in quantum cosmology lies in the emergence of a unimodular time variable \cite{Smolin_2009, Etkin:2026bwd}, the analogous notion of a bimodular time merits a dedicated Hamiltonian analysis, which would also clarify the precise propagating degrees of freedom. The Stueckelberg completion of Sec.~\ref{sec:Covariant_Reformulation} provides a natural starting point for such a treatment. But of course, we leave these questions for future work.


\acknowledgments

It brings us much pleasure in giving thanks to T.~Baker, R.~Terrazas-Santamaria, and A.~Ridley for providing helpful comments and interesting conversations---in relation to this work and in general. We are also grateful to the anonymous referee, whose detailed and constructive comments substantially improved this work. We also thank T.~Lawrence, M.~Hatteea, and H.~Wells for careful proofreading of the manuscript. J.~Hallam is supported by an STFC studentship; J.Magueijo was partly supported by STFC Consolidated Grant ST/T000791/1. For the purpose of open access, the authors have applied a Creative Commons Attribution (CC BY) licence to any Author Accepted Manuscript version arising. Supporting research data are available on reasonable request from the corresponding authors.



\bibliographystyle{JHEP}
\bibliography{biblio.bib}

@article{2015-Bufalo,
      title={How unimodular gravity theories differ from general relativity at quantum level}, 
      author={R. Bufalo and M. Oksanen and A. Tureanu},
      year={2015},
      journal={Eur. Phys. J. C},
      volume={75},
      pages={477},
      doi={https://doi.org/10.1140/epjc/s10052-015-3683-3}, 
}

@article{Riess_1998,
   title={Observational Evidence from Supernovae for an Accelerating Universe and a Cosmological Constant},
   volume={116},
   ISSN={0004-6256},
   DOI={10.1086/300499},
   number={3},
   journal={APJ},
   publisher={American Astronomical Society},
   author={Riess, Adam and Filippenko, Alexei and others},
   year={1998},
   month=sep, pages={1009–1038} }

@article{Perlmutter_1999,
  author       = {Perlmutter, Saul and Aldering, Greg and Goldhaber, Gerson and others},
  title        = {Measurements of $\Omega$ and $\Lambda$ from 42 High-Redshift Supernovae},
  journal      = {APJ},
  year         = {1999},
  volume       = {517},
  number       = {2},
  pages        = {565--586},
  doi          = {10.1086/307221}
}

@article{Brout_2022,
  author       = {Brout, Dillon and Scolnic, Dan and Popovic, Brodie and others},
  title        = {The Pantheon+ Analysis: Cosmological Constraints},
  journal      = {APJ},
  year         = {2022},
  volume       = {938},
  number       = {2},
  pages        = {110},
  doi          = {10.3847/1538-4357/ac8e04}
}

@article{Zeldovich_1968,
  author       = {Zel'dovich, Ya.},
  title        = {The Cosmological Constant and the Theory of Elementary Particles},
  journal      = {Sov. Phys. Usp.},
  year         = {1968},
  volume       = {11},
  number       = {3},
  pages        = {381--393},
  doi          = {10.1070/PU1968v011n03ABEH003927}
}

@article{Weinberg_1989,
  author       = {Weinberg, Steven},
  title        = {The Cosmological Constant Problem},
  journal      = {Rev. Mod. Phys.},
  year         = {1989},
  volume       = {61},
  number       = {1},
  pages        = {1--23},
  doi          = {10.1103/RevModPhys.61.1}
}

@article{Anderson_1971,
  author       = {Anderson, J. and Finkelstein, D.},
  title        = {Cosmological Constant and Fundamental Length},
  journal      = {Am. J. Phys.},
  year         = {1971},
  volume       = {39},
  number       = {8},
  pages        = {901--904},
  doi          = {10.1119/1.1986321}
}

@article{Alvarez_2013,
  author       = {Álvarez, Enrique and Herrero-Valea, Mario},
  title        = {Unimodular gravity with external sources},
  journal      = {JCAP},
  year         = {2013},
  volume       = {2013},
  number       = {01},
  pages        = {014},
  doi          = {10.1088/1475-7516/2013/01/014}
}

@article{Padilla_2015,
    author = "Padilla, Antonio and Saltas, Ippocratis",
    title = "{A note on classical and quantum unimodular gravity}",
    doi = "10.1140/epjc/s10052-015-3767-0",
    journal = "EPJ C",
    volume = "75",
    number = "11",
    pages = "561",
    year = "2015"
}

@article{1989-Unruh,
  title = {Unimodular theory of canonical quantum gravity},
  author = {Unruh, W.},
  journal = {Phys. Rev. D},
  volume = {40},
  issue = {4},
  pages = {1048--1052},
  year = {1989},
  month = {Aug},
  publisher = {American Physical Society},
  doi = {10.1103/PhysRevD.40.1048},
}

@article{Dragon_1988,
title = {Einstein gravity from restricted coordinate invariance},
journal = {Phys. Lett. B},
volume = {207},
number = {3},
pages = {292-294},
year = {1988},
issn = {0370-2693},
doi = {https://doi.org/10.1016/0370-2693(88)90577-1},
author = {W. Buchmüller and N. Dragon}
}

@ARTICLE{Dragon_1989,
       author = {{Buchm{\"u}ller}, W. and {Dragon}, N.},
        title = "{Gauge fixing and the cosmological constant}",
      journal = {Phys. Lett. B},
         year = 1989,
        month = jun,
       volume = {223},
       number = {3-4},
        pages = {313-317},
          doi = {10.1016/0370-2693(89)91608-0},
}

@article{1989-Henneaux,
title = {The cosmological constant and general covariance},
journal = {Phys. Lett. B},
volume = {222},
number = {2},
pages = {195-199},
year = {1989},
issn = {0370-2693},
doi = {https://doi.org/10.1016/0370-2693(89)91251-3},
author = {Marc Henneaux and Claudio Teitelboim},
}

@article{Jirousek_2023,
    author = "Jirou{\v{s}}ek, Pavel",
    title = "{Unimodular Approaches to the Cosmological Constant Problem}",
    doi = "10.3390/universe9030131",
    journal = "Universe",
    volume = "9",
    number = "3",
    pages = "131",
    year = "2023"
}

@article{Plank_2018,
   title={{Planck 2018 Results: VI. Cosmological Parameters}},
   volume={641},
   ISSN={1432-0746},
   DOI={10.1051/0004-6361/201833910},
   journal={Astronomy \& Astrophysics},
   publisher={EDP Sciences},
   author={{Planck Collaboration}},
   year={2020},
   month=sep, pages={A6} }

@article{2009-Magueijo,
    author = "Magueijo, Joao",
    title = "{Bimetric varying speed of light theories and primordial fluctuations}",
    doi = "10.1103/PhysRevD.79.043525",
    journal = "Phys. Rev. D",
    volume = "79",
    pages = "043525",
    year = "2009"
}

@article{Babichev_2008,
   title={k-Essence, superluminal propagation, causality and emergent geometry},
   volume={2008},
   ISSN={1029-8479},
   DOI={10.1088/1126-6708/2008/02/101},
   number={02},
   journal={Journal of High Energy Physics},
   publisher={Springer Science and Business Media LLC},
   author={Babichev, Eugeny and Mukhanov, Viatcheslav and Vikman, Alexander},
   year={2008},
   month=feb, pages={101–101} }

@article{BransDicke1961,
  title = {Mach's Principle and a Relativistic Theory of Gravitation},
  author = {Brans, C. and Dicke, R.},
  journal = {Phys. Rev.},
  volume = {124},
  issue = {3},
  pages = {925--935},
  numpages = {0},
  year = {1961},
  month = {Nov},
  publisher = {American Physical Society},
  doi = {10.1103/PhysRev.124.925},
}

@article{Clayton_2000,
   title={Scalar-tensor gravity theory for dynamical light velocity},
   volume={477},
   ISSN={0370-2693},
   DOI={10.1016/s0370-2693(00)00192-1},
   number={1–3},
   journal={Phys. Lett. B},
   publisher={Elsevier BV},
   author={Clayton, M.A. and Moffat, J.W.},
   year={2000},
   month=mar, pages={269–275} }

@article{Jordan_1959,
    author = {Jordan, P.},
    title = {Zum gegenwärtigen Stand der Diracschen kosmologischen},
    journal = {Hyp. Z. Phys.},
    volume = {157},
    pages = {112–121},
    year = {1959},
    doi = {10.1007/BF01375155}
}

@article{Brans_1962,
  title = {Mach's Principle and a Relativistic Theory of Gravitation. II},
  author = {Brans, C. },
  journal = {Phys. Rev.},
  volume = {125},
  issue = {6},
  pages = {2194--2201},
  numpages = {0},
  year = {1962},
  month = {Mar},
  publisher = {American Physical Society},
  doi = {10.1103/PhysRev.125.2194},
}

@ARTICLE{DESI_II,
       author = {{DESI Collaboration}},
        title = "{DESI DR2 Results II: Measurements of Baryon Acoustic Oscillations and Cosmological Constraints}",
      journal = {arXiv e-prints},
         year = 2025,
        month = mar,
          eid = {arXiv:2503.14738},
        pages = {arXiv:2503.14738},
          doi = {10.48550/arXiv.2503.14738}
}

@article{Bekenstein_1993,
  title = {Relation between physical and gravitational geometry},
  author = {Bekenstein, Jacob },
  journal = {Phys. Rev. D},
  volume = {48},
  issue = {8},
  pages = {3641--3647},
  numpages = {0},
  year = {1993},
  month = {Oct},
  publisher = {American Physical Society},
  doi = {10.1103/PhysRevD.48.3641},
}

@article{Garriga_1999,
   title={Perturbations in k-inflation},
   volume={458},
   ISSN={0370-2693},
   DOI={10.1016/s0370-2693(99)00602-4},
   number={2–3},
   journal={Phys. Lett. B},
   publisher={Elsevier BV},
   author={Garriga, Jaume and Mukhanov, V.F.},
   year={1999},
   month=jul, pages={219–225} }

@article{Gubitosi_2013EFTDE,
  author       = {Gubitosi, Giulia and Piazza, Federico and Vernizzi, Filippo},
  title        = {The Effective Field Theory of Dark Energy},
  journal      = {JCAP},
  year         = {2013},
  volume       = {2013},
  number       = {02},
  pages        = {032},
  doi          = {10.1088/1475-7516/2013/02/032}
}

@article{Koivisto_2008,
  author       = {Koivisto, Tomi },
  title        = {Disformal quintessence},
  journal      = {Phys. Rev. D},
  year         = {2008},
  volume       = {78},
  pages        = {123505},
  doi          = {10.1103/PhysRevD.78.123505}
}

@article{BraxBurrage_2014,
  author       = {Brax, Philippe and Burrage, Clare},
  title        = {Constraining Disformally Coupled Scalar Fields},
  journal      = {Phys. Rev. D},
  year         = {2014},
  volume       = {90},
  number       = {10},
  pages        = {104009},
  doi          = {10.1103/PhysRevD.90.104009}
}

@article{Sakstein_2014,
  author       = {Sakstein, Jeremy},
  title        = {Disformal Theories of Gravity: From the Solar System to Cosmology},
  journal      = {JCAP},
  year         = {2014},
  volume       = {2014},
  number       = {12},
  pages        = {012},
  doi          = {10.1088/1475-7516/2014/12/012}
}

@article{Moffat_2003,
   title={BIMETRIC GRAVITY THEORY, VARYING SPEED OF LIGHT AND THE DIMMING OF SUPERNOVAE},
   volume={12},
   ISSN={1793-6594},
   DOI={10.1142/s0218271803002366},
   number={02},
   journal={Int. J. Mod. Phys. D},
   publisher={World Scientific Pub Co Pte Lt},
   author={Moffat, J. },
   year={2003},
   month=feb, pages={281–298} }

@article{Clayton_2001,
   title={A scalar-tensor cosmological model with dynamical light velocity},
   volume={506},
   ISSN={0370-2693},
   DOI={10.1016/s0370-2693(01)00414-2},
   number={1–2},
   journal={Phys. Lett. B},
   publisher={Elsevier BV},
   author={Clayton, M.A. and Moffat, J.W.},
   year={2001},
   month=may, pages={177–186} }

@article{Magueijo_2003,
   title={New varying speed of light theories},
   volume={66},
   ISSN={1361-6633},
   DOI={10.1088/0034-4885/66/11/r04},
   number={11},
   journal={Rep. Prog. Phys.},
   publisher={IOP Publishing},
   author={Magueijo, João},
   year={2003},
   month=oct, pages={2025–2068} }

@mastersthesis{Hallam_2024,
  author       = {James Hallam},
  title        = {{Bimodular Gravity}},
  year         = {2025},
  type         = {{MSc Thesis}},
  school       = "Imperial College London",
  url          = "https://www.imperial.ac.uk/theoretical-physics/postgraduate-study/msc-in-quantum-fields-and-fundamental-forces/dissertations/"
}

@article{Chamseddine_2013,
    author = "Chamseddine, Ali H. and Mukhanov, Viatcheslav",
    title = "{Mimetic Dark Matter}",
    eprint = "1308.5410",
    archivePrefix = "arXiv",
    primaryClass = "astro-ph.CO",
    doi = "10.1007/JHEP11(2013)135",
    journal = "JHEP",
    volume = "11",
    pages = "135",
    year = "2013"
}

@article{Lim_2010,
   title={Dust of dark energy},
   volume={2010},
   ISSN={1475-7516},
   url={http://dx.doi.org/10.1088/1475-7516/2010/05/012},
   DOI={10.1088/1475-7516/2010/05/012},
   number={05},
   journal={Journal of Cosmology and Astroparticle Physics},
   publisher={IOP Publishing},
   author={Lim, Eugene A and Sawicki, Ignacy and Vikman, Alexander},
   year={2010},
   month=May, pages={012–012} }

@article{jirousek2022mimetickessence,
    author = "Jirou{\v{s}}ek, Pavel and Shimada, Keigo and Vikman, Alexander and Yamaguchi, Masahide",
    title = "{Mimetic K-essence}",
    eprint = "2212.14867",
    archivePrefix = "arXiv",
    primaryClass = "gr-qc",
    month = "12",
    year = "2022"
}

@article{Barvinsky_2021,
    author = "Barvinsky, A. O. and Kolganov, N. and Vikman, A.",
    title = "{Generalized unimodular gravity as a new form of $k$-essence}",
    eprint = "2011.06521",
    archivePrefix = "arXiv",
    primaryClass = "gr-qc",
    doi = "10.1103/PhysRevD.103.064035",
    journal = "Phys. Rev. D",
    volume = "103",
    number = "6",
    pages = "064035",
    year = "2021"
}

@book{wald:1984,
  author = {Wald, Robert M.},
  publisher = {The University of Chicago Press},
  title = {{General Relativity}},
  type = {Book},
  year = 1984
}

@article{Sawicki_2025,
   title={Causality and stability from acoustic geometry},
   volume={2025},
   ISSN={1029-8479},
   url={http://dx.doi.org/10.1007/JHEP10(2025)227},
   DOI={10.1007/jhep10(2025)227},
   number={10},
   journal={Journal of High Energy Physics},
   publisher={Springer Science and Business Media LLC},
   author={Sawicki, Ignacy and Trenkler, Georg and Vikman, Alexander},
   year={2025},
   month=Oct }

@article{Stueckelberg:1938zz,
    author = "Stueckelberg, E. C. G.",
    title = "{Interaction forces in electrodynamics and in the field theory of nuclear forces}",
    journal = "Helv. Phys. Acta",
    volume = "11",
    pages = "299--328",
    year = "1938"
}

@article{Sirera:2026klo,
    author = "Sirera, Sergi and Baker, Tessa and Hallam, James and Naidoo, Krishna",
    title = "{A Master Equation for Screening in Luminal Horndeski Gravity}",
    eprint = "2605.04154",
    archivePrefix = "arXiv",
    primaryClass = "gr-qc",
    month = "5",
    year = "2026"
}

@article{Jirou_ek_2019,
   title={{New Weyl-invariant vector-tensor theory for the cosmological constant}},
   volume={2019},
   ISSN={1475-7516},
   url={http://dx.doi.org/10.1088/1475-7516/2019/04/004},
   DOI={10.1088/1475-7516/2019/04/004},
   number={04},
   journal={Journal of Cosmology and Astroparticle Physics},
   publisher={IOP Publishing},
   author={Jiroušek, Pavel and Vikman, Alexander},
   year={2019},
   month=Apr, pages={004–004} }

@article{Chamseddine_2014,
   title={Cosmology with Mimetic Matter},
   volume={2014},
   ISSN={1475-7516},
   url={http://dx.doi.org/10.1088/1475-7516/2014/06/017},
   DOI={10.1088/1475-7516/2014/06/017},
   number={06},
   journal={Journal of Cosmology and Astroparticle Physics},
   publisher={IOP Publishing},
   author={Chamseddine, Ali H. and Mukhanov, Viatcheslav and Vikman, Alexander},
   year={2014},
   month=June, pages={017–017} }

@article{Etkin:2026bwd,
    author = "Etkin, Altay and Rassouli, Farbod-Sayyed",
    title = "{Unimodular time in JT gravity: a holographic clock}",
    eprint = "2601.07911",
    archivePrefix = "arXiv",
    primaryClass = "hep-th",
    month = "1",
    year = "2026"
}

@article{Smolin_2009,
    author = "Smolin, Lee",
    title = "{The Quantization of unimodular gravity and the cosmological constant problems}",
    eprint = "0904.4841",
    archivePrefix = "arXiv",
    primaryClass = "hep-th",
    doi = "10.1103/PhysRevD.80.084003",
    journal = "Phys. Rev. D",
    volume = "80",
    pages = "084003",
    year = "2009"
}

@book{Nakahara2003,
  title     = {Geometry, Topology and Physics},
  author    = {Nakahara, Mikio},
  edition   = {2nd},
  year      = {2003},
  publisher = {CRC Press},
  address   = {Boca Raton, FL},
  isbn      = {9780750306065},
  doi       = {10.1201/9781315275826},
  series    = {Graduate Student Series in Physics}
}


\end{document}